\documentclass[12pt]{article}

\usepackage{amssymb}
\usepackage{amsfonts}
\usepackage{amsmath}
\usepackage{bm}
\usepackage{graphicx}
\usepackage{color}
\usepackage{cite}
\usepackage{graphicx}
\usepackage{wrapfig}
\usepackage[font=small,
labelfont=bf, labelsep=period]{caption}

\definecolor{nv}{rgb}{0.1,0.1,0.6}
\definecolor{pr}{rgb}{0.2,0.1,0.5}
\definecolor{mg}{rgb}{0.4,0.0,0.4}

\newcommand{\beq}{\begin{equation}}
\newcommand{\eeq}{\end{equation}}
\newcommand{\beqy}{\begin{eqnarray}}
\newcommand{\eeqy}{\end{eqnarray}}
\newcommand{\beqyn}{\begin{eqnarray*}}
\newcommand{\eeqyn}{\end{eqnarray*}}

\newcommand{\bs}{\begin{slide}}
\newcommand{\es}{\end{slide}}
\newcommand{\bc}{\begin{center}}
\newcommand{\ec}{\end{center}}
\newcommand{\bmin}{\begin{minipage}}
\newcommand{\emin}{\end{minipage}}

\begin{document}

\begin{titlepage}
\begin{center}

{\Large\bf QCD analysis of polarized inclusive and semi-inclusive
DIS data}

\end{center}
\vskip 2cm
\begin{center}
{\bf Elliot Leader}\\
{\it Imperial College London\\ Prince Consort Road, London SW7
2BW, England }
\vskip 0.5cm
{\bf Alexander V. Sidorov}\\
{\it Bogoliubov Theoretical Laboratory\\
Joint Institute for Nuclear Research, 141980 Dubna, Russia }
\vskip 0.5cm
{\bf Dimiter B. Stamenov \\
{\it Institute for Nuclear Research and Nuclear Energy\\
Bulgarian Academy of Sciences\\
Blvd. Tsarigradsko Chaussee 72, Sofia 1784, Bulgaria }}
\end{center}

\vskip 0.3cm
\begin{abstract}
\hskip -5mm

A new combined NLO QCD analysis of the polarized inclusive and
semi-inclusive DIS data is presented. In contrast to previous
combined analyses, the $1/Q^2$ terms (kinematic - target mass
corrections, and dynamic - higher twist corrections) in the
expression for the nucleon spin structure function $g_1$ are taken
into account. The new COMPASS data are included in the analysis.
The impact of the semi-inclusive (SIDIS) data on the polarized
parton densities (PDFs)  and on the higher twist corrections is
demonstrated. The controversial behavior of the polarized strange
quark density obtained from the fit to the DIS data alone, and a
combined analysis of DIS and SIDIS data is discussed.


\end{abstract}

\end{titlepage}

\newpage
\setcounter{page}{1}

\section{Introduction}

Experiments on polarized inclusive deep inelastic lepton-hadron
scattering (DIS), reactions of the type $l+p\rightarrow l' + X $
with both polarized lepton and hadron, because of the
non-existence of neutrino data, can only, in principle yield
information on the sum of quark and antiquark parton densities
i.e. information on the polarized densities $\Delta u + \Delta
\bar{u},~\Delta d + \Delta \bar{d},~ \Delta s + \Delta
\bar{s}~\textrm{and}~ \Delta G$.

Information about the antiquark densities $\Delta \bar{u}, \Delta
\bar{d}$ and the separate $\Delta s $ and $\Delta \bar{s}$ strange
densities thus has to be extracted from other reactions, notably
polarized semi-inclusive lepton-hadron reactions (SIDIS)
$l+p\rightarrow l' +h+X$, where $h$ is a detected hadron in the
final state, or from semi-inclusive hadronic scattering (SIHS)
like $p+p\rightarrow h + X$, involving polarized protons, and only
possible at the RHIC collider at Brookhaven National Laboratory.

In contrast to the situation in unpolarized DIS, a large portion
of the most accurate data on polarized DIS lie in a kinematical
region where Target Mass Corrections (TMC) of order $M^2/Q^2$
(whose form is exactly known), and dynamical Higher Twist (HT)
corrections of order $\Lambda^2_{QCD}/Q^2$ are important
\cite{LSS07,LSS09}. We have thus included such terms in our
description of the DIS data. However, for the SIDIS data, we do
not know the analogous results at present, so do not include such
terms. As it happens almost all the SIDIS data we utilize are in
kinematic regions where such corrections should not be important.

In this talk we present the results of our combined NLO QCD
analysis of polarized inclusive and semi-inclusive DIS data
\cite{LSS10}. The new COMPASS data
\cite{{COMPASSd_h},{COMPASSd_pK},{COMPASS_A1p}} are also included
in the fit. In the calculations of the semi-inclusive asymmetries
$A_{1N}^h(x,z,Q^2)$, the NLO MRST'02 PDFs \cite{MRST02} and the
NLO DSS fragmentation functions (FFs) \cite{DSS} were used for the
unpolarized parton densities and the fragmentation functions,
respectively. The new results for the polarized PDFs are compared
to both the LSS'06 PDFs \cite{LSS07}, obtained from the fit to the
inclusive DIS data alone, and to those obtained from the first
global analysis performed by DSSV group \cite{DSSV}.

\section{Results of analysis}

In this Section we present the numerical results of our global NLO
QCD fit to the world inclusive and semi-inclusive DIS data (for
references to the data sets see our paper \cite{LSS10}). The data
used (841 experimental points for DIS and 202 experimental points
for SIDIS) cover the following kinematic regions: $\{0.005 \leq x
\leq 0.75,~~1< Q^2 \leq 62~GeV^2\}$ for DIS and $\{0.005 \leq x
\leq 0.48,~~1< Q^2 \leq 60~GeV^2\}$ for SIDIS processes. The
statistical and systematic errors are added in quadrature and the
uncertainties of the polarized PDFs presented correspond to
$\Delta \chi^2=1$. A good description of the data is achieved for
both the inclusive ($\chi^2_{NrP}$=0.85) and semi-inclusive
($\chi^2_{NrP}$=0.90) processes (NrP is the number of
corresponding experimental points). The total value of
$\chi^2_{DF}$ is 0.88. The quality of the fit to the data is
demonstrated in Fig. 1 in \cite{LSS10}.

\subsection{The role of semi-inclusive DIS data in determining
the polarized sea quark densities: Controversy about strange quark
polarization }

Due to SIDIS data a flavor decomposition of the polarized sea is
achieved and the light anti-quark polarized densities $\Delta
\bar{u}(x)$ and $\Delta \bar{d}(x)$ are determined without any
additional assumptions. While $\Delta \bar{d}(x)$ is negative for
any $x$ in the measured $x$ region, $\Delta \bar{u}(x)$ is a
positive, passes zero around $x=0.2$ and becomes negative for
large $x$. Sign-changing solutions are also found for the
polarized strange sea $\Delta \bar{s}(x)$ and gluon $\Delta G(x)$
densities. The sign-changing behavior for $\Delta G(x)$ is not
surprising since it was already found from the analysis of the
inclusive DIS data alone \cite{LSS07}. On the other hand, on the
basis of results from all published analyses of inclusive DIS, we
consider the sign-changing solution for $\Delta \bar{s}(x)$ quite
puzzling. The central values of the sea quark and gluon polarized
densities together with their error bands are presented and
compared to those of DSSV (dashed curves) in Fig. 1.
\begin{figure}[h]
\begin{center}
\resizebox{0.80\hsize}{0.45\vsize}{\includegraphics{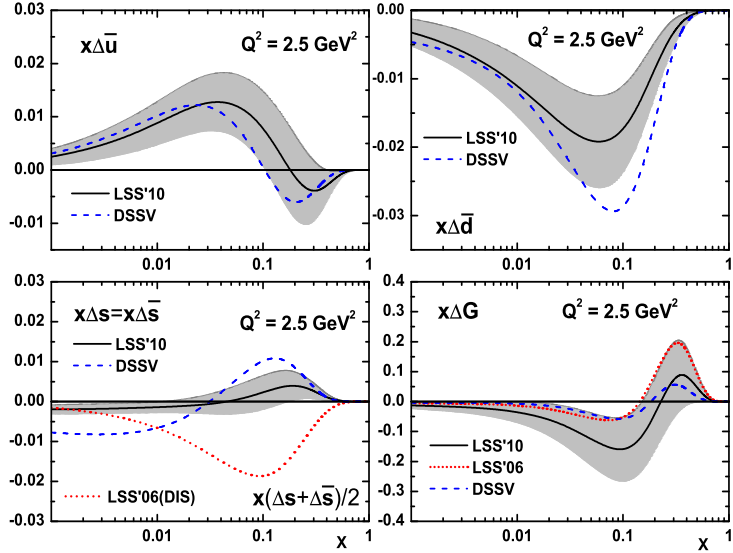}}
\caption{Our NLO sea quarks and gluon polarized PDFs at
$Q^2=2.5~GeV^2$ in the $\rm \overline{MS}$ scheme. For comparison
the DSSV PDFs \cite{DSSV} are also presented.}
\end{center}
\end{figure}

Our LSS'06 PDFs (dot curves) \cite{LSS07} obtained from the NLO
QCD analysis of the world inclusive DIS data are also presented in
Fig. 1. While the light anti-quark polarized densities $\Delta
\bar{u}(x)$ and $\Delta \bar{d}(x)$ cannot be, in principle,
determined from polarized inclusive DIS data, the sum $(\Delta s
+\Delta \bar{s})(x, Q^2)$ \emph{is} well determined and all the
NLO QCD analyses yield for this sum a {\it negative} value for any
$x$ in the measured region (for example, see Refs.
\cite{LSS07,groups}). In these analyses, however, a term like
$(1+\gamma x)$, which would permit a sign-change, was not included
in the input parametrization of $(\Delta s +\Delta \bar{s})(x,
Q^2_0)/2$. We therefore re-analysed the world polarized inclusive
DIS data using such a term in the input strange sea quark density
\begin{equation}
x(\Delta s +\Delta \bar{s})(x, Q^2_0)/2 = Ax^{\alpha}
(1-x)^{\beta}(1+\gamma x).
\end{equation}

Our preliminary results confirm the previous ones, namely, that
$(\Delta s +\Delta \bar{s})(x, Q^2)/2$ is negative in the measured
$(x,Q^2)$ region. So, the behaviour of the polarized strange quark
density remains controversial. Note that in the presence of SIDIS
data $\Delta s$ and $\Delta \bar{s}$ can, in principle, be
separately determined, as  was done recently by the COMPASS
Collaboration, where it was shown \cite{COMPASS_dels} that there
is no significant difference between $\Delta s(x)$ and $\Delta
\bar {s}(x)$ in the $x$-range covered by their inclusive and
semi-inclusive DIS data. However, the errors of the extracted
values of the difference $x(\Delta s(x)-\Delta \bar {s}(x))$ are
rather large to allow us to conclude that the assumption $\Delta
s(x)=\Delta \bar{s}(x)$ used in our's and the DSSV analyses is
correct. So, if it is not correct, it might possibly be the cause
that $(\Delta s +\Delta \bar{s})(x, Q^2)/2$ densities obtained
from the analyses of inclusive DIS data and combined inclusive and
semi-inclusive DIS data sets, respectively, are in contradiction.
However, at first glance, it looks as if the difference between
$\Delta s $ and $\Delta \bar{s}$ would have to be quite
significant and might contradict the COMPASS results. Perhaps a
more important issue is the sensitivity of the results to the form
of the fragmentation functions. An analysis by the COMPASS group
\cite{COMPASSd_pK} demonstrated that the determination of $\Delta
\bar{s}(x)$ strongly depends on the set of the fragmentation
functions used in the analysis and that the DSS FFs are crucially
responsible for the unexpected behavior of $\Delta \bar{s}(x)$
obtained from the combined analysis.
\begin{figure} [h]
\begin{center}
\resizebox{0.40\hsize}{0.23\vsize}{\includegraphics{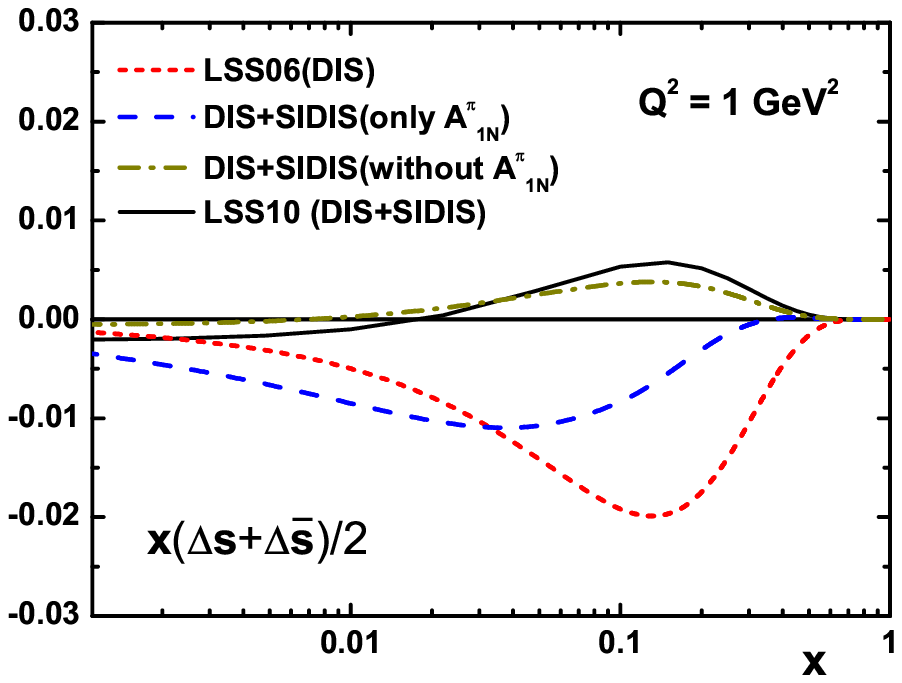}}
\resizebox{0.40\hsize}{0.23\vsize} {\includegraphics{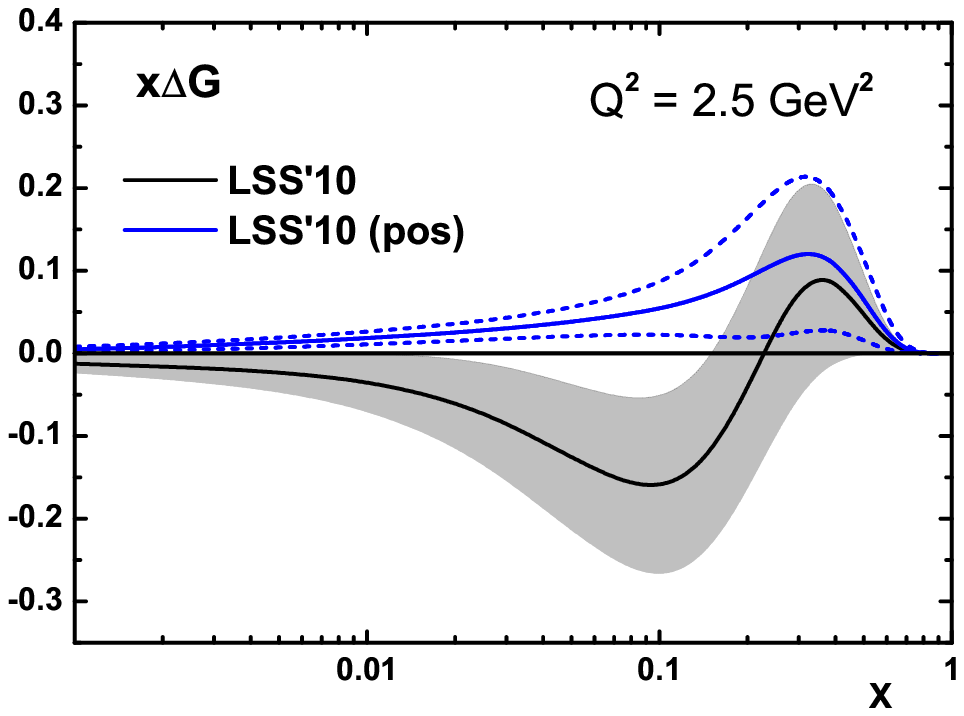}}
\caption{Dependence of the form of the polarized strange quark
density on different fits ({\bf left}). Comparison between the
positive and sign-changing gluon densities ({\bf right}).}
\end{center}
\end{figure}

In order to understand better the issue of the sign-changing
behavior of $\Delta \bar{s}(x)$ in the case of the combined fit to
the inclusive DIS and SIDIS data a more detailed analysis has been
done. In this analysis the very recent COMPASS data on the
asymmetries $A_{1,p}^{\pi_{+(-)}},~A_{1,p}^{K_{+(-)}}$ for charged
pions and kaons produced on a proton target \cite{COMPASS_dels}
have been included. First, we have performed a fit to the DIS and
SIDIS data including {\it only} the data on the pion
$A_{1N}^{\pi}$ asymmetries. Note that in this case only the sum
$x(\Delta s +\Delta \bar{s})(x, Q^2_0)$ can be determined from the
data because of the reasonable assumption
$D_s^{\pi}=D_{\bar{s}}^{\pi}$ used for all the sets of the
fragmentation functions. Second, we fitted the inclusive DIS and
SIDIS data {\it excluding} the data on the pion $A_{1N}^{\pi}$
asymmetries. The results on $x(\Delta s +\Delta \bar{s})/2$ are
illustrated in Fig. 2 (left) and compared to those obtained from
DIS (dot curve) and the combined DIS and SIDIS (solid curve)
analyses. Note that for the fragmentation functions the DSS set
was used. As seen from Fig. 2 (left), in the presence only of the
$A_{1N}^{\pi}$ data $x(\Delta s +\Delta \bar{s})/2$ (dashed curve)
is still negative in the measured $x$ region. The exclusion of
these data from the full set of SIDIS data leads to a
sign-changing behavior of $x(\Delta s +\Delta \bar{s})/2$ (dash
dot curve). Note that in the later case the assumption $\Delta
s(x)=\Delta \bar{s}(x)$ is used because the accuracy of the
present SIDIS data is not enough to separate the strange quark and
anti-quark polarized densities. One can conclude from this study
that the main reason $\Delta \bar{s}(x)$ to change a sign are the
kaon data and the kaon FFs which are less known and very different
for the different sets of fragmentation functions. So, the study
of the sensitivity of $\Delta \bar{s}(x)$ to the different kaon
FFs used in the analysis is one of the key points we plan to
investigate in the future.

In Fig. 3 we present our results for the polarized $\Delta u(x)$
and $\Delta d(x)$ densities at $Q^2=2.5~GeV^2$, which are
consistent with those obtained by DSSV (dashed blue curves). As
expected, the SIDIS data do not influence essentially the sums
$(\Delta u(x)+\Delta \bar{u}(x))$ and $(\Delta d(x)+\Delta
\bar{d}(x))$ already well determined from the analysis of the
inclusive DIS data. This fact is illustrated in Fig. 2 where our
results from the combined analysis are compared with our LSS'06
PDFs.
\begin{figure}[ht]
\begin{center}
\resizebox{0.80\hsize}{0.45\vsize}{\includegraphics{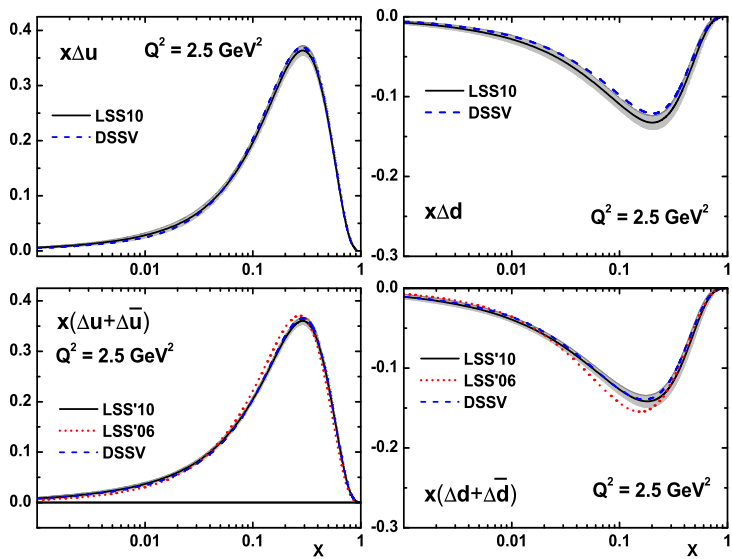}}
\caption{Our NLO $\Delta u$, $\Delta d$, $(\Delta u+\Delta
\bar{u})$ and $(\Delta d+\Delta \bar{d})$ polarized parton
densities at $Q^2=2.5~GeV^2$. DSSV \cite{DSSV} as well as LSS'06
\cite{LSS07} results for the corresponding densities are presented
too.}
\end{center}
\end{figure}

\subsection{The sign of gluon polarization}

We have found that the combined NLO QCD analysis of the present
polarized inclusive DIS and SIDIS data cannot rule out the
solution with a positive gluon polarization. The values of
$\chi^2/DF$ corresponding to the fits with sign-changing $x\Delta
G(x,Q^2)$ and positive $x\Delta G(x,Q^2)$ are practically the
same: $\chi^2/DF({\rm node}~x\Delta G )=0.883$ and
$\chi^2/DF(x\Delta G >0)=0.888$, and the data cannot distinguish
between these two solutions (see Fig. 2 (right)). The sea quark
densities obtained in the fits with positive and sign-changing
$x\Delta G(x)$ are almost identical. Note that the extracted HT
values corresponding to both fits are also effectively identical.
As a result, one can conclude that including the SIDIS data in the
QCD analysis does not help to constrain better the polarized gluon
density.
\begin{figure} [ht]
\begin{center}
\resizebox{0.40\hsize}{0.23\vsize}{\includegraphics{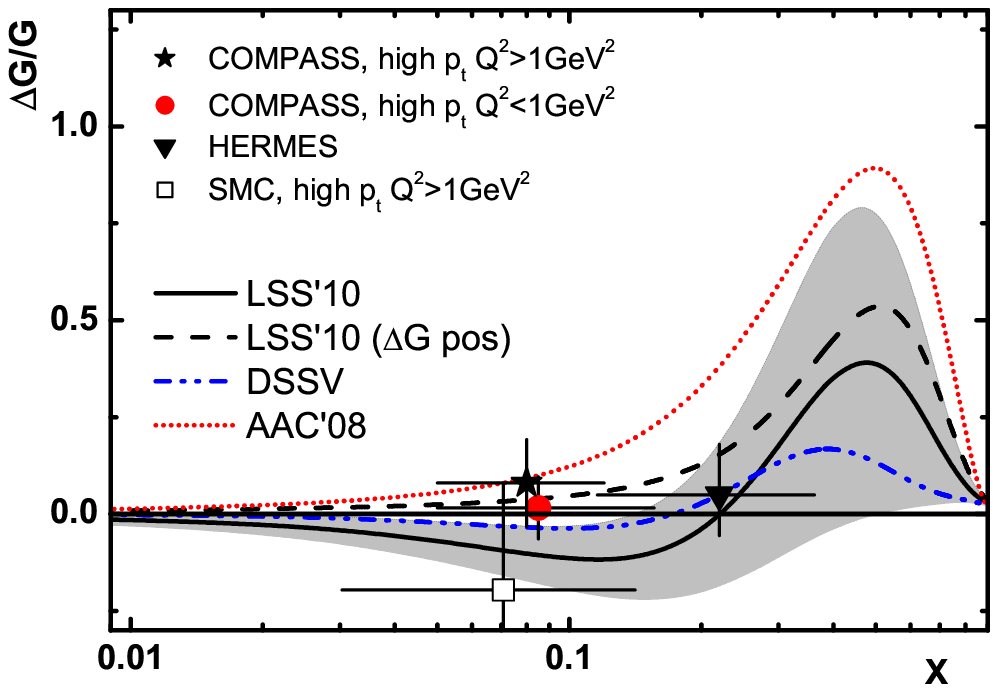}}
\resizebox{0.40\hsize}{0.23\vsize}{\includegraphics{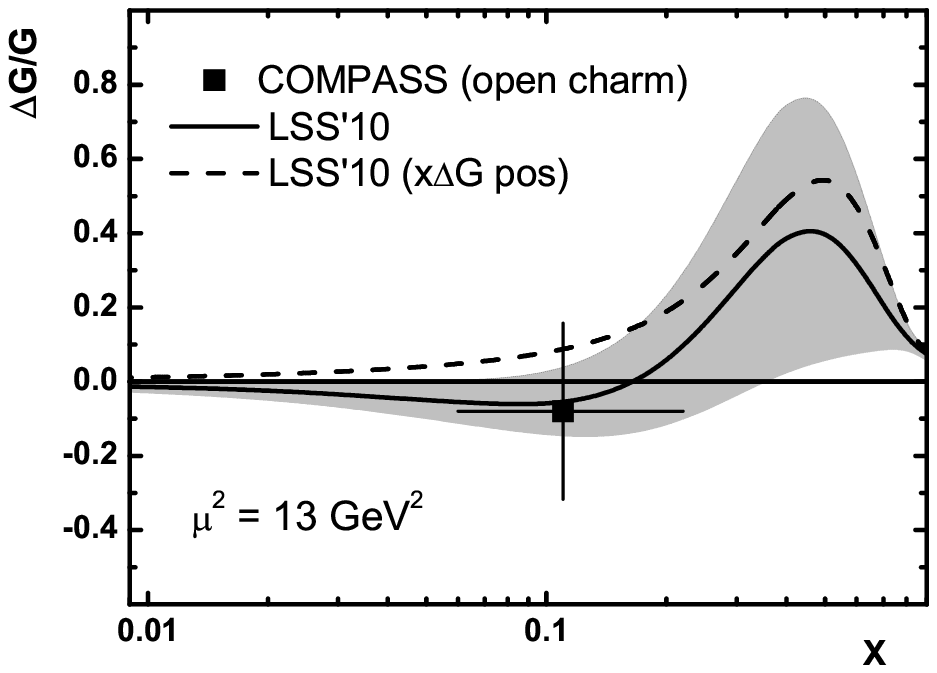}}
\caption{ Comparison between the experimental data and NLO($\rm
\overline{MS}$) curves for the ratio $\Delta G(x)/G(x)$ at
$Q^2=3~\rm GeV^2$ ({\bf left} - high $p_t$ pairs) and $Q^2=13~\rm
GeV^2$ ({\bf right} - open charm) corresponding to positive and
sign-changing $x\Delta G$. Error bars represent the total
(statistical and systematic) errors. The horizontal bar on each
point shows the $x$-range of the measurement. The NLO DSSV
\cite{DSSV} and AAC (the last Ref. in \cite{groups}) curves on
$\Delta G(x)/G(x)$ are also presented.  }
\end{center}
\end{figure}

In Fig. 4 the ratio $\Delta G(x)/G(x)$ calculated for both the
sign-changing and positive solutions for $\Delta G(x)$ obtained in
our NLO QCD analysis is compared with the directly measured values
of $\Delta G/G$ obtained from a quasi-real photoproduction of high
$p_t$ hadron pairs
\cite{SMC_high_pt,COMPASS_high_pt,HERMES_high_pt}, and from the
open charm production \cite{open_charm} measurements. For the
unpolarized gluon density $G(x)$ in the ratio above we have used
that of the NLO MRST'02 \cite{MRST02}. The theoretical curves are
given for $\mu^2=3~\rm GeV^2$ (high $p_t$ hadron pairs) and
$\mu^2=13~\rm GeV^2$ (open charm). As seen from Fig. 4,  both
solutions for the polarized gluon density are well consistent with
the experimental values of $\Delta G/G$. It should be noted,
however, that in the extraction of $\Delta G/G$ by the experiments
a LO QCD treatment has been used. A NLO extraction of the measured
values is needed in order for this comparison to be quite correct.
In conclusion, the magnitude of the gluon density $x\Delta G(x)$
obtained from our combined NLO QCD analysis of inclusive and
semi-inclusive DIS data and independently, from the photon-gluon
fusion processes, is small in the region $x\simeq 0.08-0.2$.

When our combined analysis was finished, the COMPASS Collaboration
reported the first data on the asymmetries
$A_{1,p}^{\pi_{+(-)}},~A_{1,p}^{K_{+(-)}}$ for charged pions and
kaons produced on a proton target \cite{COMPASS_dels}. As seen in
Fig. 5, our predictions for these asymmetries are in a very good
agreement with the data at measured $x$ and $Q^2$.
\begin{figure}[h]
\begin{center}
\resizebox{0.65\hsize}{0.45\vsize}{\includegraphics{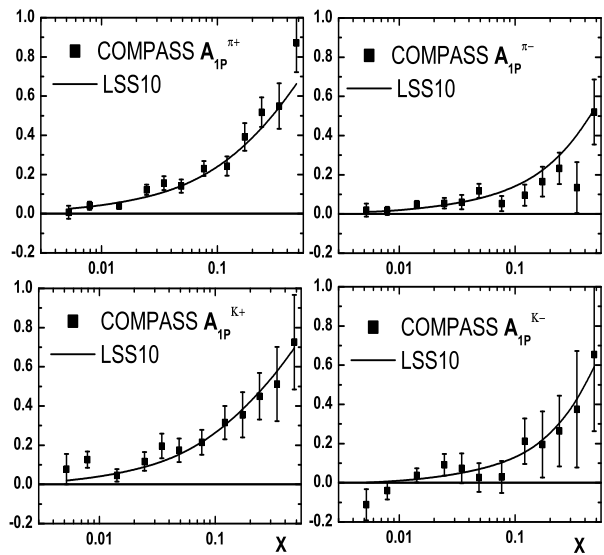}}
\caption{Our predictions for the COMPASS asymmetries for charged
 pions and kaons produced on a proton target.}
\end{center}
\end{figure}

\subsection{High twist effects}

In contrast to other combined analyses of the inclusive and
semi-inclusive DIS data, we take into account the target mass and
higher twist corrections in a the DIS sector. The values of the HT
corrections to $g_1$ extracted from the data in this analysis are
shown in Fig. 6. Compared to the HT(LSS'06) corrections obtained
in our analysis of the inclusive DIS data alone \cite{LSS07} the
values of the HT corrections for the proton target are practically
not changed, while the central values of HT corrections for the
neutron target are smaller in the region $x<0.2$, but in agreement
with $\rm HT^{(n)}$(LSS'06) within the errors, excepting the $x$
region around $x=0.1$ We consider the tendency of the $\rm
HT^{(n)}$ corrections to be smaller in the region $x<0.2$ to be a
result of the new behavior of $\Delta s(x)$ i.e. positive for $x >
0.03$. The positive contribution in $g_1^n$ from $\Delta s(x)$
should be compensated by a less positive $\rm HT^{(n)}$
contribution in this region. Since the biggest difference between
the values of $\Delta s(x)_{\rm (DIS+SIDIS)}$ and $\Delta
s(x)_{\rm DIS}$ is in the region $x\sim 0.1$ (see Fig. 1) this
effect is biggest in this $x$ region. The impact of $\Delta s(x)$
on HT corrections is visible mainly for the neutron target because
the contribution of $\Delta s(x)$ in $g_1^n$ is relatively larger
than that in $g_1^p$.
\begin{figure} [h]
\begin{center}
\resizebox{0.45\hsize}{0.35\vsize}{\includegraphics{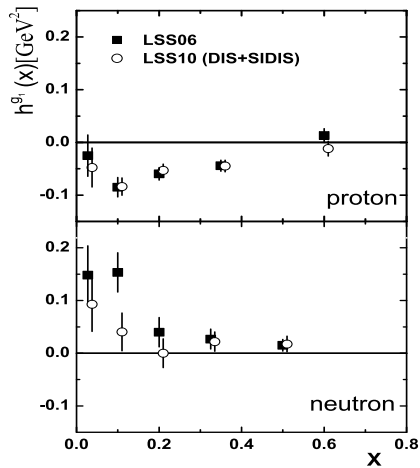}}
\caption{Impact of SIDIS data on HT effects for proton and neutron
targets. }
\end{center}
\end{figure}

Note that our results on the HT corrections to the nucleon spin
structure function $g_1(x,Q^2)$ are in a good agreement with the
phenomenological study of their first moments
\cite{h_nons_Deur,h_nons_Dubna}, as well as with the QCD sum rule
estimates \cite{Balitsky:1990jb}, the large $\rm N_c$ limit in QCD
\cite{Balla:1997hf} and the instanton model
\cite{Balla:1997hf,SidWeiss} predictions.

\subsection{The spin sum rule}

Let us finally discuss the present status of the proton spin sum
rule. Using the values for $\Delta \Sigma(Q^2)$ and $\Delta
G(Q^2)$ at $Q^2=4~GeV^2$, the first moments of the quark singlet
$\Delta \Sigma(x,Q^2)$ and gluon $\Delta G(x, Q^2)$ densities,
obtained in our analysis one  finds for the spin of the proton:
\begin{eqnarray}
J_z = \frac{1}{2}&=&\frac{1}{2}\Delta \Sigma(Q^2)+\Delta
G(Q^2)+L_z(Q^2)\nonumber\\
&=&-0.21 \pm 0.46 + L_z(Q^2)~~ ({\rm node}~\Delta G), \nonumber\\
&=&~~0.42 \pm 0.19 + L_z(Q^2)~~ ({\rm pos}~\Delta G). \label{SSR}
\end{eqnarray}
In Eq. (\ref{SSR}) $L_z(Q^2)$ is the sum of the angular orbital
momenta of the quarks and gluons. Although the central values of
the quark-gluon contribution in (\ref{SSR}) are very different in
the two cases, in view of the large uncertainty coming mainly from
the gluons, one cannot  yet come to a definite conclusion about
the contribution of the orbital angular momentum to the total spin
of the proton.

\section{Summary}

A new combined NLO QCD analysis of the polarized inclusive and
semi-inclusive DIS data is presented. In contrast to previous
combined analyses, the $1/Q^2$ terms (kinematic - target mass
corrections, and dynamic - higher twist corrections) to the
nucleon spin structure function $g_1$ are taken into account. The
new results for the  PDFs are compared to both the LSS'06 PDFs
obtained from a fit to the inclusive DIS data alone, and to those
obtained from the DSSV global analysis. The role of the
semi-inclusive data in determining the polarized sea quarks is
discussed. Due to SIDIS data $\Delta \bar{u}(x,Q^2)$ and $\Delta
\bar{d}(x,Q^2)$, as well $\Delta u(x,Q^2)$ and $\Delta d(x,Q^2)$
are determined without additional assumptions about the light sea
quarks. The SIDIS data, analysed under the assumption $ \Delta
s(x,Q^2)=   \Delta \bar{s}(x,Q^2)$, imposes a sign-changing
$\Delta \bar{s}(x,Q^2)$, as in the DSSV analysis, but our values
are smaller in magnitude, less negative at $x < 0.03$ and less
positive for $x > 0.03$. Note that $\Delta \bar{s}(x,Q^2)_{\rm
SIDIS}$ differs essentially from the negative $\frac{1}{2}(\Delta
s +\Delta \bar{s})(x, Q^2)_{\rm DIS}$ obtained from all the QCD
analyses of inclusive DIS data. It was also shown that when in the
combined QCD analysis only the pion SIDIS data are included the
polarized strange quark density is still negative for any $x$ in
the measured region, and the change-signing behavior of $\Delta
\bar{s}(x,Q^2)$ is due to the kaon asymmetries calculated by the
DSS fragmentation functions. A further detailed analysis of the
sensitivity of $\Delta \bar{s}(x,Q^2)$ to the kaon FFs is needed,
and any model independent constraints on FFs would help. Another
possible reason for this disagreement could be the assumption
$\Delta s(x,Q_0^2) =\Delta \bar{s}(x,Q_0^2)$ made in the global
analyses. However, this would probably require a significant
difference between $\Delta s $ and $\Delta \bar{s}$, which is not
seen in the COMPASS analysis. In any case, obtaining a final and
unequivocal result for $\Delta \bar{s}(x)$ remains a challenge for
further research on the internal spin structure of the nucleon.

We have found also that the polarized gluon density is still
ambiguous, and the present polarized DIS and SIDIS data cannot
distinguish between the positive and a sign-changing gluon
densities $\Delta G(x)$. This ambiguity is the main reason that
the quark-gluon contribution into the total spin of the proton is
still not well determined.

Finally, our combined NLO QCD analysis confirms our previous
results on the higher twist corrections to the nucleon spin
structure function $g_1^N$, namely, that they are not negligible
in the pre-asymptotic region and have to be accounted for in order
to extract correctly the polarized PDFs.

\vskip 0.6cm {\it Acknowledgments.} This research was supported by
the JINR-Bulgaria Collaborative Grant, by the RFBR Grants (No
08-01-00686, 09-02-01149) and by the Bulgarian National Science
Fund under Contract 02-288/2008.

\end{document}